\documentclass[3p,times]{elsarticle}

\usepackage{ecrc}

\pdfoutput=1

\volume{00}
\firstpage{1}
\usepackage{epstopdf}
\usepackage{graphicx}      % include this line if your document contains figures
\usepackage{natbib}        % required for bibliography
\usepackage{amsmath}
\usepackage{amsfonts}
\usepackage{amssymb}
\usepackage{psfrag}
\usepackage{epsfig}
\usepackage{amsfonts}
\usepackage{mathrsfs}
\usepackage[utf8]{inputenc}
\usepackage[english]{babel}
\usepackage{xparse}
\graphicspath{{figures/}}

\usepackage{arydshln}
\usepackage{color}
\newcommand{\ie}{{\it i.e. }}

\def\1{{\bf 1}_N}
\def\0{{\bf 0}}
\def\N{{\mathbb N}}
\def\R{{\mathbb R}}

 \def\G{\mathcal{G}}
 \def\V{\mathcal{V}}
 \def\L{\mathcal{L}}
 \def\P{\mathcal{P}}
  \def\C{\mathcal{C}}

\newtheorem{definition}{Definition}

\newtheorem{lemma}{Lemma}
\newtheorem{assumption}{Assumption}
\newtheorem{proposition}{Proposition}[section]
\newtheorem{remark}{Remark}
\graphicspath{{figures/}}

\usepackage{arydshln}
\usepackage{color}
\newcommand\omicron{\mathrm{o}} %vineeth: is this good?
\def\1{{\bf 1}_N}
\def\0{{\bf 0}}
\def\N{{\mathbb N}}
\def\R{{\mathbb R}}

 \def\G{\mathcal{G}}
 \def\V{\mathcal{V}}
 \def\L{\mathcal{L}}
 \def\P{\mathcal{P}}
  \def\C{\mathcal{C}}
    
\begin{document}

%\title{Tandem Cold Rolling Mill Modeling for Multi-variable Control Synthesis}
\title{Space-time budget allocation policy design  for viral marketing \tnoteref{t1}}
\tnotetext[t1]{This work has been partially funded by the CNRS PEPS project YPSOC}

%\thanks[footnoteinfo]{This work was supported by project PEPS INS2I IODINE funded by the CNRS .}

\author[CRAN]{I.C. Mor\u{a}rescu\corref{cor1}}
\ead{constantin.morarescu@univ-lorraine.fr}

\author[CRAN]{V.S. Varma}
\ead{vineeth.satheeskumar-varma@univ-lorraine.fr}

\author[UTCluj]{L. Bu\c{s}oniu},
\ead{lucian@busoniu.net}

\author[CRAN]{S. Lasaulce},
\ead{samson.lasaulce@univ-lorraine.fr}

\cortext[cor1]{Corresponding author}
\address[CRAN]{Universit\'e de Lorraine, CNRS, CRAN, F-54000 Nancy, France.}
\address[UTCluj]{Technical University of Cluj--Napoca, Romania}
%\address[L2S]{Laboratoire des Signaux et Systemes (L2S, CNRS - CentraleSupelec - Univ. Paris Sud), Gif-sur-Yvette, France.}

\begin{frontmatter}
\begin{abstract}
We address formally the problem of opinion dynamics when the agents of a social network (e.g., consumers) are not only influenced by their neighbors but also by an external influential entity referred to as a marketer. The influential entity tries to sway the overall opinion as close as possible to a desired opinion by using a specific influence budget. We assume that the exogenous influences of the entity happen during discrete-time advertising campaigns; consequently, the overall closed-loop opinion dynamics becomes a linear-impulsive (hybrid) one. The main technical issue addressed is finding how the marketer should allocate its budget over time (through marketing campaigns) and over space (among the agents) such that the agents' opinion be as close as possible to the desired opinion. Our main results show that the marketer has to prioritize certain agents over others based on their initial condition, their influence power in the social graph and the size of the cluster they belong to. The corresponding space-time allocation problem is formulated and solved for several special cases of practical interest. Valuable insights can be extracted from our analysis. For instance, for most cases, we prove that the marketer has an interest in investing most of its budget at the beginning of the process and that budget should be shared among agents according to the famous water-filling allocation rule. Numerical examples illustrate the analysis.
\end{abstract}

\begin{keyword}
Social networks, hybrid systems.
\end{keyword}

\end{frontmatter}
%===============================================================================

\section{Introduction}

%The last decades have witnessed an increasing interest in the study of opinion dynamics in social networks. This is mainly motivated by the fact that people's opinions are increasingly influenced through digital social networks. Therefore, governmental institution but also private companies consider that marketing over social networks becomes a key tool for promoting new products or ideas. However, most of the existing studies focus on the analysis of models without control, \ie they study the convergence, dynamical patterns or asymptotic configurations of the open-loop dynamics. Various mathematical models \cite{DeGroot,Friedkin,Deffuant2000,krause2002,Altafini,NilCoSa2016} have been proposed to capture more features of these complex dynamics. Empirical models based on in vitro and in vivo experiments have also been developed \cite{Davis,Ohtsubo,SamPloSOne}.\\

Opinion dynamics in social networks has become a problem of increasing research interest during the last decades. This can be explained by the multiplication of digital social networks that allow a faster and more persistent influence of opinions. In this context, governmental institution and private companies use marketing over social networks as a key tool for promoting their products or ideas. However, to the best of our knowledge, there is no formal analysis pointing out the improvements that can be achieved by using the network topology in the design of the marketing strategy. Indeed, most of the existing studies focus on the analysis of models without control, \ie they study the convergence, dynamical patterns or asymptotic configurations of the open-loop opinion dynamics. Various mathematical models \cite{DeGroot,Friedkin,Deffuant2000,krause2002,Altafini,NilCoSa2016} have been proposed to capture different features of these complex dynamics. Empirical models based on in vitro and in vivo experiments have also been developed \cite{Davis,Ohtsubo,SamPloSOne}.

One controversial problem is related to emergence of consensus in social networks. Social studies pointed out that, in general, opinions tend to converge one toward another during interactions. Therefore, is not surprising that consensus received a particular attention in opinion dynamics literature \cite{Axelrod1997,GalamMoscovici1991}. While some mathematical models naturally lead to consensus \cite{DeGroot,Friedkin}, others lead to network clustering \cite{krause2002,Altafini,MG10}. In order to enforce consensus, some recent studies propose the control of one or a few agents, see \cite{Camponigro,Dietrich}. Besides these methods of controlling opinion dynamics towards consensus, we also find recent attempts to control the discrete-time dynamics of opinions such that as many agents as possible reach a certain set after a finite number of influences \cite{Hegselmann}. Another relatively new line of research is based on the change of opinions through the change of susceptibility and resistance parameters \cite{Fogg2002,MariekeEtAl2015,Abebe2018}. Basically, each individual is characterized by certain parameters that make it more or less easy to influence. In \cite{Abebe2018} (and some reference therein) the authors zoom in the model and see how the persuasion can be realized by acting on the susceptibility of individuals. Instead, we are looking directly at the outcome of the persuasion strategy and use this information in the long term evolution of opinion dynamics. 

Viral marketing refers to the practice where a seller attempts to artificially create word-of-mouth advertising among potential customers, and the effectiveness of this trend has been well established by social scientists and economists \cite{leskovec2007dynamics,arthur2009pricing}. In \cite{masucci2014strategic}, the authors consider multiple influential entities competing to control the opinion of consumers under a game theoretical setting. However, this work assumes an undirected graph and a voter model for opinion dynamics resulting in strategies that are independent of the node centrality.  On the other hand, \cite{varma2017opinion} considers a similar competition with opinion dynamics over a directed graph and no budget constraints.\\

In this paper, we consider a different problem that requires minimizing the distance between opinions and the desired value using a given control/marketing budget. Moreover, we assume that the maximal marketing influence cannot instantaneously shift the opinion of one individual to the desired value. Basically, we consider a continuous time opinion dynamics and we want to design a marketing strategy that minimizes the distance between opinions and the desired value after a given finite number of discrete-time campaigns under budget constraints.The main motivation for this choice is the time scale of relevant events. The campaign refers to sales before some events and their duration is much smaller than the duration of the spreading of opinions related to the advertised products. There exist many practical situations where the use of an hybrid OD model seems completely natural.  For instance, during a presidential campaign it is common to measure the opinions of the electors through polls just before, and just after, a time-localized event such as a big political meeting or a TV debate (see e.g., \cite{mcclurg2006electoral}). Despite its natural relevance, to the best of the authors’ knowledge, no hybrid controlled OD model has been proposed to study the opinion dynamics in social networks under an external influence. To solve this control design problem we write the overall closed-loop dynamics as a linear-impulsive system and we show that the optimal strategy is to influence as much as possible the most central/popular individuals (see \cite{Bonacich} for a formal definition of  
centrality) of the network as far as the graph modeling the social network is weakly connected (\ie it contains at least a directed spanning tree). We also point out that the budget allocation has to take into account the size of clusters (maximal subsets of weakly connected agents, see \cite{MG10}) when the graph is disconnected.  \\
%The problem considered in this paper is very relevant for the field of economics and marketing. In order to improve the existing marketing strategies  \citep{friedman1958game,esmaeili2009game,edelman2007internet} we assume that the social network topology is known. Therefore, instead of handling the network as a homogeneous population our advertising strategy detects and targets the most popular/influential members.\\

To the best of our knowledge, our work is different from all the existing results on opinion dynamics control. Unlike the few previous works on the control of opinions in social networks, we do not control the state of the influencing entity. Instead, we consider that value as fixed and we control the influence weight that the marketer has on different individuals of the social network. By doing so, we emphasize the advantages of targeted marketing with respect to broadcasting (uniform) strategies when budget constraints have to be taken into account. Moreover, we show that, although the individual control action $u_i(t_k)$ at time $t_k$ can be chosen in the interval $[0,\bar{u}]$, the optimal choice is discrete: either $0$ or $\bar{u}$.\\

The rest of the paper is organized as follows. Section \ref{sec:model} formulates the opinion dynamics control problem under consideration. A useful preliminary result for solving a specific optimization problem with constraints is given in Section \ref{prelim}. To motivate our analysis, we emphasize in Section \ref{ex_mot} the improvements that can be obtained by targeted advertising with respect to a uniform/broadcasting control. Section \ref{main} contains the results related to the optimal control strategy. We first analyze the case when the campaign budget is given a priori and must be optimally partitioned among the network agents. Secondly, we look at the case when the campaign budget is unknown but the campaigns are distanced in time. Thirdly, we consider the case of large networks that can be approximated as a union of clusters/sub-networks. All these three cases point out that the optimal control contains only $0$ or $\bar{u}$ actions. These results motivate us to study in Section \ref{Sec:discrete_action} the space-time distribution of the budget under the assumption that all the components of $u(t_k)$ are either $0$ or $\bar{u}$. We conclude that the budget has to be allotted according to the influence power of each agent which in turn depends on the initial condition, centrality and the size of the clusters in which it lies. Numerical examples and concluding remarks end the paper.

\section{Problem statement}
\label{sec:model}

We consider an entity (for example, a company) that is interested in attracting consumers to some product (electrical cars, healthy food, etc). Consumers belong to a social network and we refer to any consumer as an agent. For the sake of simplicity, we consider a fixed social network over the set of vertices $\mathcal{V}= \{1,2,\dots,N\}$ of $N$ agents. In other words, we identify each agent with its index in the set $\mathcal{V}$. To agent $i \in \mathcal{V}$ we assign a normalized scalar opinion $x_i(t) \in [0,1]$ that evolves under the influence of neighbors' opinions and external entity persuasion/advertising action. We use $x(t) = (x_1(t),x_2(t),\dots,x_N(t))^\top$ to denote the state of the network at any time $t$, where $x(t) \in \mathcal{X}$ and $\mathcal{X}= [0,1]^N$.

In order to obtain a larger market share with a minimum investment, the external entity applies an action vector on marketing campaigns at discrete time instants. The set of campaigns time instants is finite: $\mathcal{T}=\{t_0,t_1,\dots,t_M\}$. The number of campaigns $M$ is considered to be finite but arbitrarily large because we are interested in the finite (arbitrarily large) time behavior of the network.  A given action therefore corresponds to a given marketing campaign aiming at influencing the consumer's opinion. Between two consecutive campaigns, the consumer's opinion is only influenced by the other consumers of the networks. We assume that $t_{k}-t_{k-1}=\delta_k\in[\delta_{\min},\delta_{\max}]$ where $0<\delta_{\min}<\delta_{\max}$ are two fixed real numbers.

Throughout the paper we consider $d\in\{0,1\}$ be the desired opinion that the external entity would like to be adopted for all the consumers. We also consider $\forall i\in\V$ the following dynamics:
\begin{equation}\label{eq_dynamics}
\left\{\begin{split}
&\dot{x}_i(t)=\sum_{j=1}^Na_{ij} [x_j(t)-x_i(t)], \quad  t\in[t_k,t_{k+1})\\
&x_i(t_k)=u_i(t_k)d+[1-u_i(t_k)] x_i(t_k^-)
\end{split}\right.,\  \forall k\in\N,
\end{equation}
where $u_i(t_k)\in[0,\bar{u}],\ \forall i\in\V$, where $\bar{u} \in (0,1)$ is a saturation on each component of the control, and \begin{equation}
\sum_{k=0}^M \sum_{i=1}^N u_i(t_k)\le B
\end{equation}
where $B$ represents the total budget of the external entity for the marketing campaigns.

It is worth pointing out that external influences are modeled through a sequence of impulsive dynamics (second equation of \eqref{eq_dynamics}). This corresponds either to the case when the duration of the campaign is much shorter than the time between two consecutive campaigns or to the case in which the real dynamics during the campaign is neglected and only the resulting state is used as an entry for the next inter-campaign period. It can also be noticed that the state-jump resulting from external influence is both related to the budget allocated to Agent $i$ at time $t_k$ i.e., $u_i(t_k)$ and the value of the state before advertising $x_i(t_k^-)$. While the former proportionality is intuitive the later expresses an increasing resistance of individuals while approaching the advertised state.\\
It is also important to highlight that we assume a uniform behavior of the agents with respect to external influence. In real social networks, some agents (central ones for instance) may be harder to influence. This means that for a given value of the external influence their state jump will be smaller than the jump of other agents under the same external influence. This can be done by adding a scaling factor in the second equation of \eqref{eq_dynamics}.

Dynamics \eqref{eq_dynamics} can be rewritten using the collective variable $X(t)=[d, x(t)^\top]^\top$ as:

\begin{equation}\label{eq_collective_dynamics}
\left\{\begin{split}
&\dot{X}(t)=-\L X(t)\\
&X(t_k)=\P X(t_k^-)
\end{split}\right. ,
\end{equation} where

\[\L=\left(\begin{array}{cc} 0 & \0_{1,N}\\ \0_{N,1} & L\end{array}\right),\ \P=\left(\begin{array}{cc} 1 & \0_{1,N}\\ u(t_k) & I_N-\mathrm{diag}(u(t_k))\end{array}\right)\]

with $\mathrm{diag}(u(t_k))\in\R^{N\times N}$ being the diagonal matrix having the components of $u(t_k)$ on the diagonal. Here, $L$ is the Laplacian matrix associated to the graph formed by the adjacency matrix elements $a_{i,j}$,  i.e., $L_{ij} = -a_{ij}$ for $i \neq j$ and $L_{ii} = \sum_{i \neq j} a_{ij}$.
\begin{definition}
The (vector) centrality of Agent $i$ is the $i^{th}$ component of the left eigenvector $v$ of $L$ associated with the eigenvalue $0$ and satisfying $v^\top\1=1$.
\end{definition}
\begin{remark}It is worth noticing that:
\begin{itemize}
\item $\L$ is a Laplacian matrix corresponding to a network of $N+1$ agents. The first agent represents the external entity and is not connected to any other agent while the rest of the agents represents the consumers and interact through the social network defined by the influence weights $a_{ij}$.
\item $\P$ is a row stochastic matrix that can be interpreted as a Perron matrix associated with the directed graph having the external entity as a central node. This node is not influenced by the network and keeps its state constant. On the other hand it possibly influences (notice that components of $u(t_k)$ can be 0) all the other nodes. Consequently, without budget constraints, the network can reach, at least asymptotically, the state $d \1$.
\end{itemize}
\end{remark}
Several space-time control strategies can be implemented under budget constraints. For instance, we can spend the same budget for each agent \ie $u_i(t_k)=u_j(t_k),\ \forall (i,j) \in \V^2$, we can also allocate the entire budget for specific agents of the network. Moreover, the budget can be spent either on few or many campaigns.
Our objective is to design a space-time control strategy that minimizes the following cost function
%\begin{equation}
%J^T=\sum_{i=1}^N |x_i(T)-d|
%\end{equation}
%for some $T>t_M$, and we have the cost associated with the asymptotic opinion given by
\begin{equation}
J^{\infty}=\sum_{i=1}^N \lim_{t\rightarrow\infty}|x_i(t)-d|
\end{equation}
This can be interpreted as follows. If the entity (a company for example) is interested in convincing the public to buy some product or change their habits (practice sports or quit smoking for instance), it will try to move the asymptotic consensus value of the network as close as possible from the desired value, i.e. minimize $J^\infty$. In some other cases, like an election campaign which targets to get the opinions close to $d$ within a finite time $T$, we will minimize $J^T=\sum_{i=1}^N |x_i(T)-d|$. Therefore, in the absence of additional campaigns, the agents will exchange their opinion through the network and asymptotically converge to certain local or global agreements. It is worth noting that, after the last campaign, the system state converges exponentially fast. Indeed, in the absence of campaigns, the system dynamics is just $\dot{x}(t)=-Lx(t)$ which has the consensus manifold as a global uniformly exponentially stable attractor. This means that $x_i(T)$ is a good approximation of $\lim_{t\rightarrow\infty}x_i(t)$ when $T$ is sufficiently large.

\section{Preliminaries}\label{prelim}

Before starting the analysis of the multi-agent system in the presence of external influence, we state a key lemma which will help us to find the optimal solutions for the considered scenarios for the budget allocation problem.
\begin{lemma}\label{lem1}
Given an optimization problem (OP) under the following standard form
\begin{equation}
\begin{array}{cl}
 \underset{y \in \mathbb{R}^N}{\text{minimize}} &  C(y)\\
\text{subject to}  &  y_i - \bar{y} \leq 0, \ \forall i \in  \{1, ...,N\} \\
                           & - y_i \leq 0, \ \forall i \in  \{1, ...,N\}  \\
                          & \displaystyle{\sum_{i=1}^N} y_i - B \leq 0
\end{array} \label{eq:stdOP}
\end{equation}

where $N \in \mathbb{N}, N \geq 1$, $\bar{y}<1$, $B \geq 0$ and $C(y)$ is a decreasing convex function in $y_i$ such that one of the following two conditions hold.\\
\textbf{Case 1:} $\forall \ i \in \{1,\dots,N\},\ \exists g(y) \geq 0$ such that $$ \displaystyle\frac{\partial C(y)}{\partial y_i} = -c_i g(y)$$\\[1mm]
for some $c_i \in \mathbb{R}$.\\
\textbf{Case 2:} $ \displaystyle\frac{\partial C(y)}{\partial y_i} =- \frac{1}{1-y_i}$ for all $i \in \{1,\dots,N\}$.\\[1mm]
Then an optimal solution $y^*$ to this OP is given by water-filling as follows
\begin{equation}
y_{\omicron(i)}^* = \left\{ \begin{array}{lll}
\bar{y} & \text{if} & i \leq \left\lfloor \frac{B}{\bar{y}}\right \rfloor  \vspace{.1cm} \\
B- \bar{y}  \left\lfloor \frac{B}{\bar{y}}\right\rfloor   & \text{if} & i = \left\lfloor \frac{B}{\bar{y}}\right\rfloor +1  \\
0 & \text{otherwise} &
\end{array}  \right. \label{eq:gensol}
\end{equation}
where $\omicron: \{1,\dots,N \} \mapsto  \{1,\dots,N\}$ represents an ordering function which can be any bijection for Case 2 and, one satisfying $ c_{\omicron(1)} \geq c_{\omicron(2)} \geq \dots \geq c_{\omicron(N)}$ for Case 1.
\end{lemma}
{\bf Proof:}
Note that all the constraint functions of the considered OP are affine, which corresponds to sufficient conditions for applying KKT conditions. Since the OP is convex, KKT conditions are necessary and sufficient for optimality. By denoting the Lagrangian by 
\begin{equation}
\ell (y, \lambda, \lambda',  \widehat{\lambda}) = C(y) +  \sum_{i=1}^N \lambda_i ( y_i - \bar{y} ) - \sum_{i=1}^N \lambda_i' y_i + \widehat{\lambda} ( \displaystyle{\sum_{i=1}^N} y_i - B ),
\end{equation}
the first necessary and sufficient condition for optimality writes:
\begin{equation}
-\nabla C(y^\star) = \displaystyle{\sum_{i=1}^N \lambda_i^{\star} \nabla (y_i^\star-\bar{y}) - \sum_{i=1}^N (\lambda_i^\star)' \nabla y_i^\star}
+ \widehat{\lambda}^\star \nabla \left(\displaystyle \sum_{i=1}^N y_i^\star - B \right).
 \label{eq:KKT1}
\end{equation}
The primal feasibility conditions write
\[ 0 \leq y_i^\star \leq \bar{y} \, \, \forall i \in \{1,\dots,N\} \]
and \begin{equation}
\sum_{i=1}^N y_i^\star \leq B.  \label{eq:sumconst}
\end{equation}
All the KKT multipliers must satisfy the dual feasibility conditions: $\lambda_i^\star \geq 0$, $(\lambda_i^\star)' \geq 0$, $ \widehat{\lambda}^\star \geq 0$ for all $i\in \{1,\dots,N\}$. At last, the complementary slackness conditions are given by
\[
\hspace{-2cm}\begin{split}
&\lambda_i^\star (y_i^\star-\bar{y}) =0, \\
&(\lambda_i^\star)' y_i^\star =0 , \\
&\widehat{\lambda}^\star \left[ \left(\sum_{i=1}^N y_i^\star \right) - B \right] =0.   \end{split}\]
\textbf{Case 1:} Let us assume that $\displaystyle\frac{\partial C(y)}{\partial y_i} = c_i g(y)$. Then, we can simplify (\ref{eq:KKT1}) for component $i$ to get
\begin{equation}
 c_i g(y^\star)  = \widehat{\lambda} + \lambda_i^\star -  (\lambda_i^\star)'
\end{equation} 
which must hold for all $i \in \{1,\dots,N\}$. Since $g(y^\star)$ is identical for all $i \in \{ 1, \dots,N \}$ and is non-negative, we must have $\lambda_i^\star$, $ (\lambda_i^\star)'$ and $\widehat{\lambda}$ chose so that the above equation holds. Take $y^\star$ from (\ref{eq:gensol}). Set $\lambda_j^\star=(\lambda_j^\star)'=0$ for $j= \left(\omicron\left\lceil \frac{\beta_k}{\bar{u}}\right\rceil \right) $ as it is the only component with a non-saturated solution. For any $i$ such that $\omicron(i) < j$, we have $c_i \geq c_j$ and this can be satisfied by setting $y_i^\star=\bar{u}$ and having $\lambda_i^\star>0$ and $\lambda'^\star_i=0$. On the other hand, for any $i$ such that $\omicron(i) > j$, we set $y^*_i=0$ and the KKT conditions are satisfied if $\lambda_i^\star=0$ and $(\lambda_i^\star)' >0$. The solution from (\ref{eq:gensol}) can also be verified to satisfy (\ref{eq:sumconst}) and therefore, we have it satisfying all the KKT conditions.\\
\color{black}
\textbf{Case 2:} When $\displaystyle\frac{\partial C(y)}{\partial y_i} =\frac{-1}{1-y_i}$, the role of $c_i$ is replaced by $\displaystyle\frac{-1}{1-y_i}$ and the agent with $y_i^\star=0$ has the lowest absolute value on the left side of the Lagrangian (\ie $1$), and the agent with the saturation $y_i^\star=\bar{y}$ has the highest absolute value (i.e., $\displaystyle\frac{1}{1- \bar{u}}$). Due to symmetry, any agents can be chosen to have the min or max saturation.  \hfill $\blacksquare$

Lemma 1 provides a water-filling type optimal allocation policy for the problem considered. The solution in Case 1 is to select the best agent in terms of the coefficient $c_n$, allocate the maximum possible $y_n$ to it, then allocate the remaining budget to the next best agent and so on. This implies sorting the agents based on $c_n$, which is done using the function $\omicron$, and saturating the $y_n$ for the first $ \left\lfloor \frac{B}{\bar{y}}\right\rfloor$ agents, assigning the remaining budget to the next agent and $0$ to the rest.

\section{Performance analysis of the considered benchmark strategy}\label{ex_mot}

As a reference strategy, we consider the broadcasting-based marketing. For every campaign, it consists in allocating the available campaign budget uniformly among all the consumers. However, the campaign budget is allowed to vary over time that is, from campaign to campaign, under the total budget constraint. In order to highlight the potential of designing more advanced strategies, we show here that for some particular network topologies it is possible to quantify analytically the potential gain brought by implementing target marketing (\ie using space-time strategies) over broadcasting strategies.

First, let us compute the optimal revenue that we can get by broadcasting strategies  \ie $u_i(t_k)=u_j(t_k)\triangleq \alpha_k,\ \forall i,j\in \V$. We suppose that the graph representing the social network contains a directed spanning tree (\ie a directed graph in which, except the root which is not influenced, each node is influenced by a single other node called parent). Let $v$ be the left eigenvector of $L$ associated with the eigenvalue 0 and satisfying $v^\top \1=1$. Therefore, in the absence of any control action, one has that $\lim_{t \to \infty} x(t) = v^\top x(0)\1\triangleq  {x}^{\infty}_0$. Let us also introduce the following notation:
\begin{equation}
 {x}^{\infty}_k=\lim_{t\rightarrow\infty}e^{-L(t-t_k)}x(t_k)=v^\top x(t_k)\1,\quad \forall k\in\N.
\end{equation}

Following \eqref{eq_collective_dynamics} and using $\delta_k=t_{k+1}-t_k,\ D_k=diag(u(t_{k}))$ one deduces that:

\begin{equation}
\begin{split}
 {x}^{\infty}_{k+1}=v^\top x(t_{k+1})\1=v^\top\left[u(t_{k+1})d+(I_N-D_{k+1})x(t_{k+1}^-)\right]\1
=v^\top\left[u(t_{k+1})d+(I_N-D_{k+1})e^{-L\delta_k}x(t_k)\right]\1.
\end{split}
\end{equation}

Since $v^\top L=\0_N$ one has that $v^\top e^{-L\delta_k}=v^\top$ and consequently one obtains that

\begin{equation}\label{recurrence1}
 {x}^{\infty}_{k+1}- {x}^{\infty}_k=v^\top\left(u(t_{k+1})d-D_{k+1}e^{-L\delta_k}x(t_k)\right)\1.
\end{equation}

In the case of broadcasting one has $u(t_k)=\alpha_k \1$ and $D_k=\alpha_k I_N$, where $\alpha_k \in [0,\bar{u}]$ for all $k \in \{0,\dots,M\}$. Therefore, using $v^\top\1=1$, \eqref{recurrence1} becomes

\begin{equation}
 {x}^{\infty}_{k+1}- {x}^{\infty}_k=\alpha_{k+1}(d\1- {x}^{\infty}_k),
\end{equation}

which can be equivalently rewritten as

\begin{equation}\label{recurrence2}
(d\1- {x}^{\infty}_{k+1})=(1-\alpha_{k+1})(d\1- {x}^{\infty}_k).
\end{equation}

Using \eqref{recurrence2} recursively one obtains that
\begin{equation}
J^{\infty}(\alpha)  =|\1^\top(d\1- {x}^{\infty}_{M})|=\prod_{\ell=0}^{M}(1-\alpha_\ell)|\1^\top(d\1- {x}^{\infty}_0)|
= \left(N d -   \1^\top  {x}^{\infty}_0 \right) \prod_{\ell=0}^{M}(1-\alpha_\ell).
\end{equation}
where $J^\infty(\alpha)$ denotes the cost associated with a broadcasting strategy using $\alpha_k$ at campaign $t_k$.
\begin{proposition}
The cost $J^{\infty}(\alpha)  $ obtained when implementing broadcasting is minimized by using the maximum possible investments as soon as possible, i.e., for all $k \in \{0,\dots,M\}$,
\begin{equation}
\alpha_k  = \left\{ \begin{array}{lll}
\bar{u} & \text{if} & k \leq \left\lfloor \frac{B}{N\bar{u}}\right \rfloor  \vspace{.1cm} \\
\frac{B}{N}- \bar{u}  \left\lfloor \frac{B}{N\bar{u}}\right\rfloor   & \text{if} & k = \left\lfloor \frac{B}{N\bar{u}}\right\rfloor +1  \\
0 & \text{otherwise} &
\end{array}  \right.\label{eq:solvbrod}
\end{equation}
\end{proposition}
{\bf Proof:}
Minimizing $J^{\infty}(\alpha)$ under broadcasting strategy assumption is equivalent with the minimization of $\prod_{\ell=0}^{k+1}(1-\alpha_\ell)$. This is equivalent to minimizing
\begin{equation}
C(\alpha) =\log\left(\prod_{\ell=0}^{k+1}(1-\alpha_\ell) \right)
\end{equation}
and we have
\begin{equation}
\frac{\partial C}{\partial \alpha_\ell} = -\frac{1}{1-\alpha_{\ell}}
\end{equation}
This results in an OP which satisfies the conditions to use Lemma 1 Case 2. \hfill $\blacksquare$

It is noteworthy that for $u_i\in[0,1)$ one has that
\begin{equation}\label{ineqB}
\prod_{\ell=0}^{k+1}(1-\alpha_\ell)\ge 1-\sum_{\ell=0}^{k+1}\alpha_\ell\ge 1-\frac{B}{N}.
\end{equation} The last inequality in \eqref{ineqB} comes from the broadcasting hypothesis $u_i(t_\ell)=\alpha_\ell,\ \forall i\in\V$ and consequently the budget spent in the $\ell-$th campaign is $N\cdot \alpha_\ell$. Therefore, the total budget for $k+2$ campaigns is $N\sum_{\ell=0}^{k+1}\alpha_\ell$ and has to be smaller than $B$. \\
Thus

\begin{equation}
J=|\1^\top(d\1- {x}^{\infty}_{k+1})|\ge (1-\frac{B}{N})|\1^\top(d\1- {x}^{\infty}_0)|.
\end{equation}

The interpretation of \eqref{ineqB} is that for broadcasting strategy the minimal cost $J$ is obtained when the whole budget is spent in one marketing campaign (provided this is possible \ie $B\leq N\bar{u}$), otherwise the first inequality in \eqref{ineqB} becomes strict meaning that

\begin{equation}
J>  (1-\frac{B}{N})|\1^\top(d\1- {x}^{\infty}_0)|.
\end{equation}

Let us now suppose that the graph under consideration is a directed spanning tree having the first node as root. Then, using a targeted marketing in which the external entity influences only the root, we will show that, under the same budget constraints, the cost $J$ will be smaller. Indeed, for this graph topology one has $v=(1,0,\ldots,0)^\top$ yielding $ {x}^{\infty}_k=x_1(t_k)\1$. Moreover, the dynamics of $x_1(\cdot)$ writes as:

\begin{equation}\label{eq_dynamics_x1}
\left\{\begin{split}
&\dot{x}_1(t)=0, \qquad  t\in[t_k,t_{k+1})\\
&x_1(t_k)=u_1(t_k)d+(1-u_1(t_k))x_1(t_k^-)
\end{split}\right.,\  \forall k\in\N.
\end{equation}

Therefore,
\begin{equation}
x_1(t_k)=u_1(t_k)d+(1-u_1(t_k))x_1(t_{k-1})
\end{equation}

yielding
\begin{equation}
d-x_1(t_k)= [1-u_1(t_k)][d-x_1(t_{k-1})],
\end{equation}

which is equivalent to \eqref{recurrence2}. As we have seen before, in the broadcasting strategies one has $\sum_{\ell=0}^{k+1}\alpha_\ell\le\frac{B}{N}$ while targeting only the root, the constraint becomes $\sum_{\ell=0}^{k+1}u_1(t_\ell)\le B$. Therefore, for any given broadcasting strategy $(u_1,u_2,\ldots,u_k)$ there exists a targeted on the root strategy that consists in repeating $N$ times $(u_1,u_2,\ldots,u_k)$. Doing so, one obtains
\begin{equation}
(d\1- {x}^{\infty}_{k+1})=\left[\prod_{\ell=0}^{k+1}(1-\alpha_\ell)\right]^N(d\1- {x}^{\infty}_0).
\end{equation}
which leads to a cost which is seen to be less than the one obtained when using broadcasting-based marketing. 

\section{General optimal control strategy}\label{main}
First, we rewrite the optimal control problem as an optimization problem by treating the control $u(t_k)$ as an $NM-$dimensional vector to optimize. We denote $u_{i,k}=u_i(t_k)$ to represent the control for agent $i \in \mathcal{V}$ at time $t_k$. Then our problem can be rewritten as

\begin{equation}
\begin{array}{cl}
\underset{u \in \mathbb{R}^{NM}}{\text{minimize}} & \ J^\infty(u) \\
\text{subject to} & u_{i,k} - \bar{u} \leq 0, \ \ \ \forall i \in \mathcal{V}, k \in \{0,\dots,M\}\\
& - u_{i,k} \leq 0,  \ \ \ \forall i \in \mathcal{V}, k \in \{0,\dots,M\}\\
&  \displaystyle{\sum_{i=1}^N \sum_{k=1}^M} u_{i,k} - B \leq 0
\end{array} \label{eq:genOP}
\end{equation}

%\begin{equation}
%\begin{array}{c}
%\underset{u \in \mathbb{R}^{NM}}{\text{Minimize}} \ J^\infty(u) \\[3mm]
%\text{Subject to } 0 \leq u_{i,k} \leq \bar{u} \ \ \ \forall i \in \mathcal{V}, k \in \{0,\dots,M\} ,\\
%\text{ and } \sum_{i=1}^N \sum_{k=1}^M u_{i,k} \leq B
%\end{array} \label{eq:genOP}
%\end{equation}

Here, $J^\infty(u)$ is seen as a multilinear function. Before solving problem \eqref{eq:genOP} we want to get further insights on structure of the optimal solution, which will lead to important simplifications.
Therefore, instead of solving the general optimization problem \eqref{eq:genOP}, we first consider splitting our problem into time-allocation and space-allocation.

\begin{assumption}\label{ass1}
The graph $\G=(\V,L)$ is weakly connected (sometimes referred to as quasi-strongly connected) \ie it contains at least one directed spanning tree.
\end{assumption}
Assumption \ref{ass1} is standard in the analysis of multi-agent systems and guarantees that information flows over the entire network. In our analysis this assumption is not essential but we start analyzing networks that satisfy it and, in a second step, we solve the budget allocation problem over disconnected networks. When Assumption \ref{ass1} holds, if we know that for campaign $k$ a maximum budget of $\beta_k \leq B$ has been allocated (\ie, for a given time-allocation), we find the optimal control strategy for the $k-$th campaign. Moreover, for long campaign duration (\ie $t_{k+1}-t_k$ large) and  given time budget allocation $(\beta_0,\dots,\beta_M)$, we provide a computationally oriented optimal space allocation of the budget. Based on these results, we propose a discrete-action space-time control strategy. Next, we extend the results to the case when Assumption \ref{ass1} does not hold and the network consists of a union of weakly connected clusters.

\subsection{Minimizing the per-campaign cost}
In this section we consider that Assumption \ref{ass1} holds and the budget $\beta_k$ for each campaign is a priori given. The objective is to find the spatial allocation of the budget that optimizes the cost $|\1^\top(d\1- {x}^{\infty}_k)|$ associated with the time allocation $(\beta_0,\dots,\beta_M)$. Consequently, the following  budget constraint has to be considered at campaign $k$:
\begin{equation}
\sum_{i=1}^N u_i(t_k) \leq \beta_k.
\label{eq:stagebudgetcons}
\end{equation}
The associated cost for the campaign $k$ is written as
\begin{equation}
\begin{array}{l}
J^{\infty}_k(u(t_k))  =|\1^\top(d\1- {x}^{\infty}_k)|=N\cdot|d - \displaystyle\sum_{i=1}^N v_i x_i(t_k)| 
 =N\cdot \left|d - \displaystyle\sum_{i=1}^N v_i (u_i(t_k) d + [1-u_i(t_k)] x_i(t_k^-) )  \right|.
\end{array}
\end{equation}
This rewriting of the cost allows us to define the right quantity to measure the influence power of an agent, which translates the gain the marketer can make by investing on this agent. The corresponding quantity is defined and used in the next proposition.

\begin{proposition}\label{prop:connected} Define the influence power of Agent $i$ as 
$p^k_i =  v_i |d -x_i(t_k^-)|$.  Denote by $\pi_k: \mathcal{V} \to  \mathcal{V}$, a bijection which sorts the agents based on decreasing $p^k_i$, \ie $p^k_{\pi_k(1)}\geq p^k_{\pi_k(2)}\geq \dots \geq p^k_{\pi_k(N)}$. Under Assumption \ref{ass1} the cost $J^{\infty}_k(u(t_k))$ is minimized by the following investment profile
\begin{equation}
u_{\pi(i)}^* (k)= \left\{ \begin{array}{lll}
\bar{u} & \text{if} & i \leq \left\lfloor \frac{\beta_k}{\bar{u}}\right \rfloor  \vspace{.1cm} \\
\beta_k - \bar{u}  \left\lfloor \frac{\beta_k}{\bar{u}}\right\rfloor   & \text{if} & i = \left\lfloor \frac{\beta_k}{\bar{u}}\right\rfloor +1  \\
0 & \text{otherwise} &
\end{array}  \right. \label{eq:solvbrod}
\end{equation}.
\end{proposition}
{\bf Proof:}
Note that minimizing $J^{\infty}_k(u(t_j))$ is equivalent with the minimization of
\[ C(u(t_k))= \left( d - \sum_{i=1}^N v_i (u_i(t_k) d + (1-u_i(t_k))x_i(t_k^-) )  \right)^2  \]
with the constraints $0 \leq u_i(t_k) \leq \bar{u}$ for all $i \in \mathcal{V}$ and (\ref{eq:stagebudgetcons}). We notice that
\begin{equation}
\frac{\partial C}{ \partial u_i(t_k)} = - 2 v_i (d-x_i(t_k^-)) (d- x_k^{\infty})
\end{equation} where we used the notation $x_k^{\infty}=v^\top x(t_k)$.\\
If $d =1$, then $x_i(t_k^-) \leq 1,\ \forall i\in\V$ and $$(d-x_i(t_k^-)) (d- x_k^{\infty})\ge0.$$ On the other hand, if $d=0$, we have $x_i(t_k^-) \geq 0,\ \forall i\in\V$ and $$(d-x_i(t_k^-)) (d- x_k^{\infty})\ge0.$$
Therefore, we can rewrite the above equation as
\begin{equation}\label{eq:KKTfinal}
\frac{\partial C}{ \partial u_i(t_k)} = -\gamma_i g(u(t_k)),\qquad g(u(t_k))=|d- x_k^{\infty}|
\end{equation}
which satisfies the conditions to use Case 1 of Lemma \ref{lem1}. \hfill $\blacksquare$

\subsection{Space allocation for long campaign duration}

In the following we consider that Assumption \ref{ass1} holds and a finite number of marketing campaigns with a priori fixed budget are scheduled such that $t_{k+1}-t_k$ is very large for each $k \in \{0,1,\dots,M-1\}$. Due to Assumption \ref{ass1} and the long duration of the campaigns, we can assume that $x_i(t_{k+1}^-) = x_k^{\infty}$ for all $i \in \mathcal{V}$ and $k \in \{0,1,\dots,M-1\}$. Under this assumption, we write
\begin{equation}
x_i(t_1^-) = x_0^{\infty}(u(t_0))= \sum_{i=1}^N v_i (d u_i(t_0)+  x_i(t_0^-) (1-u_i(t_k)) )
\end{equation}
for any $i \in \mathcal{V}$. Subsequently, we have
\begin{equation}\label{longstagedynamics}
x_k^{\infty}(u(t_0),u(t_1),\dots,u(t_k))= \displaystyle\sum_{i=1}^N v_i \left[ d u_i(t_k)+ x_{k-1}^{\infty}(u(t_0),\dots,u(t_{k-1}))(1-u_i(t_k))  \right]
\end{equation}
for all $k \in \{1,2,\dots,M\}$. Our objective is to minimize $$J^\infty(u)=N\cdot \big|x_M^{\infty}(u(t_0),\dots,u(t_M)) -d\big|$$ and this can be done using the Proposition \ref{prop:longstage}  below. 

\begin{proposition}\label{prop:longstage} Define $\rho_k : \mathcal{V} \to \mathcal{V}$ a bijection such that $\rho_0 = \pi_0$ (defined in Proposition \ref{prop:connected}) and for all $k \in \{1,2,\dots,M\}$, $\rho_k$ gives the agent index after sorting over $v_i$, \ie $v_{\rho_k(1)} \geq v_{\rho_k(2)} \geq \dots \geq v_{\rho_k(N)}$. Let Assumption \ref{ass1} hold and the time budget allocation be given by $\beta= (\beta_0,\dots,\beta_M)$ such that $\displaystyle\sum_{k=1}^M \beta_k \leq B$ and $\beta_k \leq N \bar{u}$. Then, the optimal allocation per agent minimizing the cost $J(u)$ is given by
\begin{equation}
u_{\rho_k(i)}^* (k)= \left\{ \begin{array}{lll}
\bar{u} & \text{if} & i \leq \left\lfloor \frac{\beta_k}{\bar{u}}\right \rfloor  \vspace{.1cm} \\
\beta_k - \bar{u}  \left\lfloor \frac{\beta_k}{\bar{u}}\right\rfloor   & \text{if} & i = \left\lfloor \frac{\beta_k}{\bar{u}}\right\rfloor +1  \\
0 & \text{otherwise} &
\end{array}  \right. \label{eq:solvbrod}
\end{equation}.
\end{proposition}

{\bf Proof:}
Minimizing $J^\infty(u)$ is equivalent with the minimization of
\[ C(u)= \left( d - x_M^{\infty}(u(t_0),\dots,u(t_M)) \right)^2  \]
with the constraints given in (\ref{eq:genOP}).

We have
\[
\frac{\partial C(u)}{\partial u_{i,k}} = 2 (d- x_M^{\infty}) \frac{\partial x_M^{\infty}}{\partial u_{i,k}}
\]
for any $i \in \mathcal{V},\ k \in \{0,1,\dots,M\}$.
Observe that
\[
\frac{\partial x_k^{\infty}}{\partial u_{i,k}} = v_i ( d - x_{k-1}^{\infty} )
\]
for $k\in \{1,\dots,M\}$ and
\[
\frac{\partial x_0^{\infty}}{\partial u_{i,0}} = \pm \gamma_i
\]
with the sign being negative if $x_i(t_0^-) >d$ and positive otherwise. We also have
\[
\frac{\partial x_k^{\infty}}{\partial u_{i,k-1}} = \frac{\partial x_{k-1}^{\infty}}{\partial u_{i,k-1}} \sum_{i \in \mathcal{V}} v_i (1-u_{i,k})
\]
Using the equations above iteratively, we have
\[
\frac{\partial x_M^{\infty}}{\partial u_{i,k}} = v_i ( d - x_{k-1}^{\infty} )  \prod_{j=k+1}^M \sum_{i \in \mathcal{V}} v_i (1-u_{i,j})
\]
for $k \geq 1$ and
\[
\frac{\partial x_M^{\infty}}{\partial u_{i,0}} = \pm \gamma_i \prod_{j=1}^M \sum_{i \in \mathcal{V}} v_i (1-u_{i,j})
\]
Therefore,
\[
\frac{\partial C(u)}{\partial u_{i,k}} = 2 (d- x_M^{\infty}) v_i ( d - x_{k-1}^{\infty}(u) )  \prod_{j=k+1}^M \sum_{i \in \mathcal{V}} v_i (1-u_{i,j})
\]
for $k \geq 1$ and
\[
\frac{\partial C(u)}{\partial u_{i,0}} = 2 |d- x_M^{\infty}(u)| \gamma_i \prod_{j=1}^M \sum_{i \in \mathcal{V}} v_i (1-u_{i,j})
\]

Assume that the optimal time-allocation of budget is known and is given by $\beta=(\beta_0,\beta_1,\dots,\beta_M)$ such that
\[
\sum_{i=1}^N u_{i,k} = \beta_k,\quad \forall k \in \{0,1,\dots,M\}
\]
Then, the optimal spatial allocation problem within any campaign $k$ is an OP satisfying Case 1 of Lemma \ref{lem1}. \hfill $\blacksquare$

Due to the long stage duration, the opinions are in consensus for all $t_k$, except for the case of $k=0$, which is the first campaign. This means that in the first stage, the agents are sorted based on both their initial opinions and their centralities, but from the next stage onwards, only their centralities are considered.

\subsection{Budget allocation for clusterized networks}\label{sec:cluster}

In practice, when the social network becomes large, several issues may appear. Computational complexity limitations may prevent from operating with $NM-$dimensional spaces. Also, uncertainties or inaccuracies on $L$ or the centrality vector may appear. Motivated by these two observations, we formulate here the solution in the case where the network is structured into clusters which may be naturally present or arise from a dynamical process. Clusters can be determined from cluster detection algorithms such as the one proposed in \cite{blondel2008,MG10}) and emphasize a time-scale separation in the dynamics of the system. The subsequent analysis could use the time-scale modeling (\cite{ChowKokotovic,Arcak2007,MMN-Time-scale2016}) but we would have to carefully deal with the jumps introduced by the advertising campaigns. Instead of doing that, in this paper we simplify the analysis and the presentation by neglecting the weakest interconnections that may appear between agents belonging to different clusters. In other words we analyze the behavior of the reduced-order dynamics. Therefore, the network is assumed to be the union of a certain number of weakly connected clusters; in that case, we show that the cluster size has to be taken into account to optimally allocate the available budget. 

Let us consider that $\V=\bigcup_{i=1}^m \C_i$, where $\C_1,\ldots,C_m$ are disjoint subsets of agents which are weakly connected. For any $i\in\{1,\ldots,m\}$ we also denote by $N_i$ the cardinality of cluster $\C_i$ and by $x_{C_i}\in\R^{N_i}$ the column vector collecting all the states of the agents in cluster $\C_i$. Since Assumption \ref{ass1} does not hold, only local agreements corresponding to each cluster are obtained. Basically, $L=\mathrm{diag}(L_1,\ldots,L_m)$ where $L_i\in\R^{N_i\times N_i}$ is the Laplacian matrix associated with the interactions in cluster $\C_i$. In the sequel we denote by $x_{\C_i,k}^\infty$ the agreement value of cluster $\C_i$ starting from the initial condition $x_{C_i}(t_k)$. In other words, if the last advertising campaigns takes place at time $t_k$ than the system converges to the following state:
\begin{equation}
 {x}_k^\infty=\lim_{t\rightarrow\infty}e^{-L(t-t_k)}x(t_k)=\left(\begin{array}{c}x_{\C_1,k}^\infty{\bf 1}_{N_i}\\ \vdots\\x_{\C_m,k}^\infty{\bf 1}_{N_m}\end{array}\right),\quad \forall k\in\N.
\end{equation}

A fixed number of advertising campaigns is assumed and the last campaign takes place at time $t_M$. The overall cost to minimize can be expressed as:
\begin{equation}
J^\infty(u)=\sum_{i=1}^mN_i\cdot \big|x_{C_i,M}^{\infty}(u(t_0),\dots,u(t_M)) -d\big|. \end{equation}

In order to characterize the optimal control strategy in the case of disconnected networks we introduce some additional notation: $v^i$ is the left eigenvector of $L_i$ associated with the simple eigenvalue 0 and satisfying $(v^i)^\top{\bf{1}_{N_i}}=1$. It can noticed that 
\begin{equation}\label{longstagedynamicscluster}
x_{C_i,k}^{\infty}(u(t_0),u(t_1),\dots,u(t_k))= \displaystyle\sum_{j\in\C_i} v^i_j \left[ d u_j(t_k)+ x_{\C_i,k-1}^{\infty}(u(t_0),\dots,u(t_{k-1}))(1-u_j(t_k))  \right]
\end{equation}
for all $k \in \{1,2,\dots,M\}$. This observation is exploited in the next proposition to define an appropriate measure of the influence power of an agent. Here again, the corresponding quantity is used to express the optimal budget allocation policy.

\begin{proposition}\label{prop:cluster} Let $\V=\bigcup_{i=1}^m \C_i$, where $\C_1,\ldots,C_m$ are disjoint subsets of agents which are weakly connected. Let also the time budget allocation be given by $\beta= (\beta_0,\dots,\beta_M)$ such that $\displaystyle\sum_{k=1}^M \beta_k \leq B$ and $\beta_k \leq N \bar{u}$. For each agent $j\in\V\cap\C_i$ define the influence power as $s_j^k=N_i\cdot v^i_j\cdot|d-x_j(t_k^-)|$. At last, define $\sigma_k:\V\mapsto\V$ a bijection which sorts the agents based on decreasing $s_j^k$ \ie $s^k_{\mathcal{H}(1)} \geq s^k_{\mathcal{H}(2)} \geq \dots \geq s^k_{\mathcal{H}(N)}$. Then, the optimal allocation per agent minimizing the cost $J(u)$ is given by
\begin{equation}
u_{\sigma_k(i)}^* (k)= \left\{ \begin{array}{lll}
\bar{u} & \text{if} & i \leq \left\lfloor \frac{\beta_k}{\bar{u}}\right \rfloor  \vspace{.1cm} \\
\beta_k - \bar{u}  \left\lfloor \frac{\beta_k}{\bar{u}}\right\rfloor   & \text{if} & i = \left\lfloor \frac{\beta_k}{\bar{u}}\right\rfloor +1  \\
0 & \text{otherwise} &
\end{array}  \right. \label{eq:cluster}
\end{equation}
\end{proposition}
{\bf Proof:}
The result follows again by applying Case 1 of Lemma \ref{lem1}. We avoid unnecessary details and we just point out that
\begin{equation}
\frac{\partial J(u)}{\partial u_{j,k}} =-s^k_j
\end{equation} leading to the desired result when $g$ is identically 1 in Lemma \ref{lem1}.
\hfill $\blacksquare$

\section{Discrete-action space-time control strategy}\label{Sec:discrete_action}

Motivated by the results in Propositions \ref{prop:connected} and \ref{prop:longstage}, in this section we consider that $u_i(t_k)\in\{0,\bar{u}\}, \forall i\in\V, k\in\N$ and $B= Q \bar{u}$ with $Q\in\N$ given a priori. The objective is to numerically find the best space-time control strategy for a given initial state $x_0$ of the network.

% \subsection{Algorithms}

\subsection{Brute force algorithm}

We will consider in turn the cases of short and long campaigns. 

In the short-campaign case, given a time allocation consisting of the budgets $\beta_k = b_k \bar{u}$ at each campaign, either Proposition \ref{prop:connected} (if the directed graph is weakly connected) or Proposition \ref{prop:cluster} (if the network is clustered) tells us how to allocate each campaign budget optimally across the agents. Denote all possible budgets at one campaign by $\mathcal{B} = \{0, \dotsc, \min\{N, Q\}\}$. A simple algorithm is then to search in a brute-force manner all possible time allocations $\boldsymbol{b} = (b_0, \dotsc, b_M) \in \mathcal{B}^{M+1}$, subject to the constraint $\sum_k b_k \leq Q$. For each such vector $\boldsymbol{b}$, we simulate the system from $x_0$ with dynamics \eqref{eq_dynamics} where the budget $b_k$ is allocated with Proposition \ref{prop:connected} or \ref{prop:cluster}. After the last campaign, we compute with the appropriate formula the final, infinite-time state of the network $x_F(\boldsymbol{b})$. We retain a solution with the best average cost:
$$
\min_{\boldsymbol{b}} \frac{1}{N} \sum_{i=1}^N \vert x_{i,F}(\boldsymbol{b}) - d \vert
$$
where subscript $i$ is the agent index (recall that the agents all have the same opinion at infinite time if the network is weakly connected, and if it is clustered each cluster has its own opinion). Note that we report the average cost over the agents, $J^\infty/N$, instead of the sum $J^\infty$ because this version is easier to interpret as a mean deviation of each agent from the target state. Furthermore, the simulation can be done in closed form, using the fact that $x^-(t_{k+1}) = e^{-L \delta_k} x(t_k)$. The complexity of this search is $O(N^3 (M+1) (\min\{N, Q\}+1)^{M+1})$, dominated by the exponential term. Therefore, this approach will only be feasible for small values of $N$ or $Q$, and especially of $M$.

Considering now the long-campaign case for weakly connected networks, we could still implement a similar brute-force search, but using dynamics \eqref{longstagedynamics} for inter-campaign propagation and Proposition \ref{prop:longstage} for allocation over agents. However, now we can do better by taking advantage of the fact that for all $k > 1$, the opinions of all the agents reach identical values. Using this, we will derive a more efficient, dynamic programming solution to the optimal control problem:
$$
\min_{\boldsymbol{b}} \vert x_{1,F}(\boldsymbol{b}) - d \vert
$$
where the long-campaign dynamics apply but by a slight abuse of notation we keep it the same as above.

\subsection{Dynamic programming algorithm}

When the graph is weakly connected and we are in the long-campaign case, we are able to provide a dynamic programming (DP) algorithm that is much more efficient than the brute force search above. Owing to the long campaigns, the agents have already reached consensus by $t_1$ and so we can use a scalar variable $y_k=|d-x^{\infty}_{k-1}|$ to represent the cost (or equivalently, the state of the network) before the campaign for all $k \in \{1,2,\dots,M+1\}$, i.e., from the second campaign onwards. After the initial-campaign decision at $k=0$, $y_1$ is computed in a special way since the network is not yet at consensus. To see how, consider a fixed, given initial opinion $x(t_0)$. Then, $y_1=f_0(b_0)$ where $f_0$ describes the evolution of the network after allocating the first-campaign budget $b_0$ using Proposition \ref{prop:connected}:
\begin{equation}
f_0(b_0)=\sum_{i=1}^{b_0} v_{\rho_0(i)} (1-\bar{u}) |x_{\rho_0(i)}(t_0) -d| +  \sum_{i=b_0+1}^N v_{\rho_0(i)} |x_{\rho_0(i)}(t_0) -d| \label{eq:cost0stage}
\end{equation}

Now, to compute $y_k$ after the decisions at subsequent campaigns $k \in  \{1,2,\dots,M\}$, \ie for $t_1,t_2,\dots,t_M$, define function $f: \mathbb{Z}_{\geq 0} \to (0,1]$:
\begin{equation}
f(b) =  \left(1-\bar{u} \sum_{i=1}^{b}  v_{\rho_1(i)} \right).
\end{equation}
If $b_k$ denotes the budget allocated to campaign $k$, then it can be shown that after the optimal spatial allocation described in Proposition \ref{prop:longstage},
\begin{equation}
y_{k+1}=y_k f(b_k),
\end{equation}
for all $k \in  \{1,2,\dots,M\}$. This lets us write the final cost as
\begin{equation}
y_{M+1} = y_1 \prod_{k=1}^M f(b_k).
\end{equation}

In addition to the network state, we will need an additional integer state $r_k \in \{0, \dotsc, Q\}$ that keeps track of the remaining budget. This variable is initialized to the total budget $r_0=Q$ and evolves according to $r_{k+1} = r_k - b_k$. 

For any given $y_1$ obtained after using a budget $b_0$ during campaign $0$, minimization of the final cost involves minimizing $\prod_{k=1}^M f(b_k)$, with the constraint $\sum_{k=1}^M b_k \leq r_1$. Since $f(b) >0$ for any $b \geq 0$, we can minimize the final cost by minimizing the logarithm of the product mentioned above, i.e., the minimization of the final cost is equivalent to the following optimization problem:
\begin{equation}\label{eq:logproblem}
\begin{array}{c}
\min_{b_0,b_1,\dots,b_M} \log(f_0(b_0)) +  \sum_{k=1}^M \log(f(b_k)),\\
\text{such that } \sum_{k=0}^M b_k \leq Q ,\\
\text{where } b_k \in \{0,1,\dots,\min\{Q,N\}\}, \ \forall k \in \{0,\dots,M\}
\end{array}
\end{equation}
where the budget allocated to each campaign is upper-bounded by the budget available and the number of agents in the network as assumed in Proposition \ref{prop:cluster}.

In order to implement the DP algorithm, we keep a value function $V_k$ that represents at each $k$ the sum of logarithms from step $k$ onwards. This value function only depends on the remaining budget:
\begin{equation}
\begin{array}{l}
V_M(r_M)= \log(f(r_M)), \\
V_k(r_k)= \min_{b_k \in \{0,1,\dots,\min\{N,r_k\} \}} \left[] \log(f(b_M))  +V_{k+1}(r_k-b_k) \right]  \,\,\,\, \text{for } k \in \{M-1, M-2,\dots,1\},\\
V_0 = \min_{b_0 \in \{0,1,\dots,\min\{N,Q\} \}}  \left[\log(f_0(b_0)) +V_{1}(Q-b_0)\right]
\end{array}
\end{equation}
To understand this algorithm, note first that because $f$ is a strictly decreasing function, the optimal budget to use at campaign $M$ is all the remaining budget, which leads to the final-campaign DP initialization $V_M(r_M)=\log(f(r_M))$. Then, at intermediate campaigns, we simply minimize the summation in \eqref{eq:logproblem} with a DP rule. Finally, for the initial campaign $k=0$, instead of using $f$, the cost for any initial budget $b_0$ is calculated explicitly by taking $\log(f_0(b_0))$ defined in \eqref{eq:cost0stage}.  Note that $V_0$ is a scalar constant, that represents the total optimal cost of the solution.

Once $V_k$ is available, an optimal solution is found by a forward pass, as follows:
\begin{equation}
    \begin{array}{ll}
b_0^*  = \arg\min_{b_0 \in \{ 0, 1, \dots, \min\{Q, N\}\}}\left[  \log(f_0(b_0)) +V_{1}(Q-b_0) \right]\\
b_k^*  = \arg\min_{b_k \in \{0, 1, \dots, \min\{r_k, N\}\}}  \left[\log(f(b_k)) +V_{k+1}(r_k-b_k)\right], \quad \text{for } k=1, \dotsc, M-1\\
b_M^*  = r_M\\
\end{array}
\end{equation}

The complexity of the backward pass for value function computation is $O
\left(MNQ\min\{N, Q\}\right)$ (the complexity of the forward pass is much smaller). To develop an intuition, take the case $N < Q$; then the algorithm is quadratic in $N$ and linear in $M$ and $Q$. This allows us to apply the algorithm to much larger problems than the brute-force search above. 
% Finally, note that in principle we could develop a DP algorithm for the short-campaign problem, but there we cannot condense the network state into a single number and so we cannot derive the efficient decomposition of the cost . Each agent state would have to be discretized instead, leading to a memory and time complexity proportional to $Y^N$, which makes the algorithm unfeasible for more than a few agents.

\color{black}

\subsection{Numerical results}

In this section, we begin by exemplifying on a small-scale problem: the short-campaign brute-force algorithm; long-campaign DP; and the clustered case (again with the brute-force method). After that, to illustrate the scalability of the proposed DP method with respect to the size of the network and to the number of campaigns, we devise a larger experiment with more agents and campaigns, and apply DP to it.

\begin{figure}[!htb]
  \vspace*{1em}
  \centering
  \includegraphics[width=0.4\textwidth]{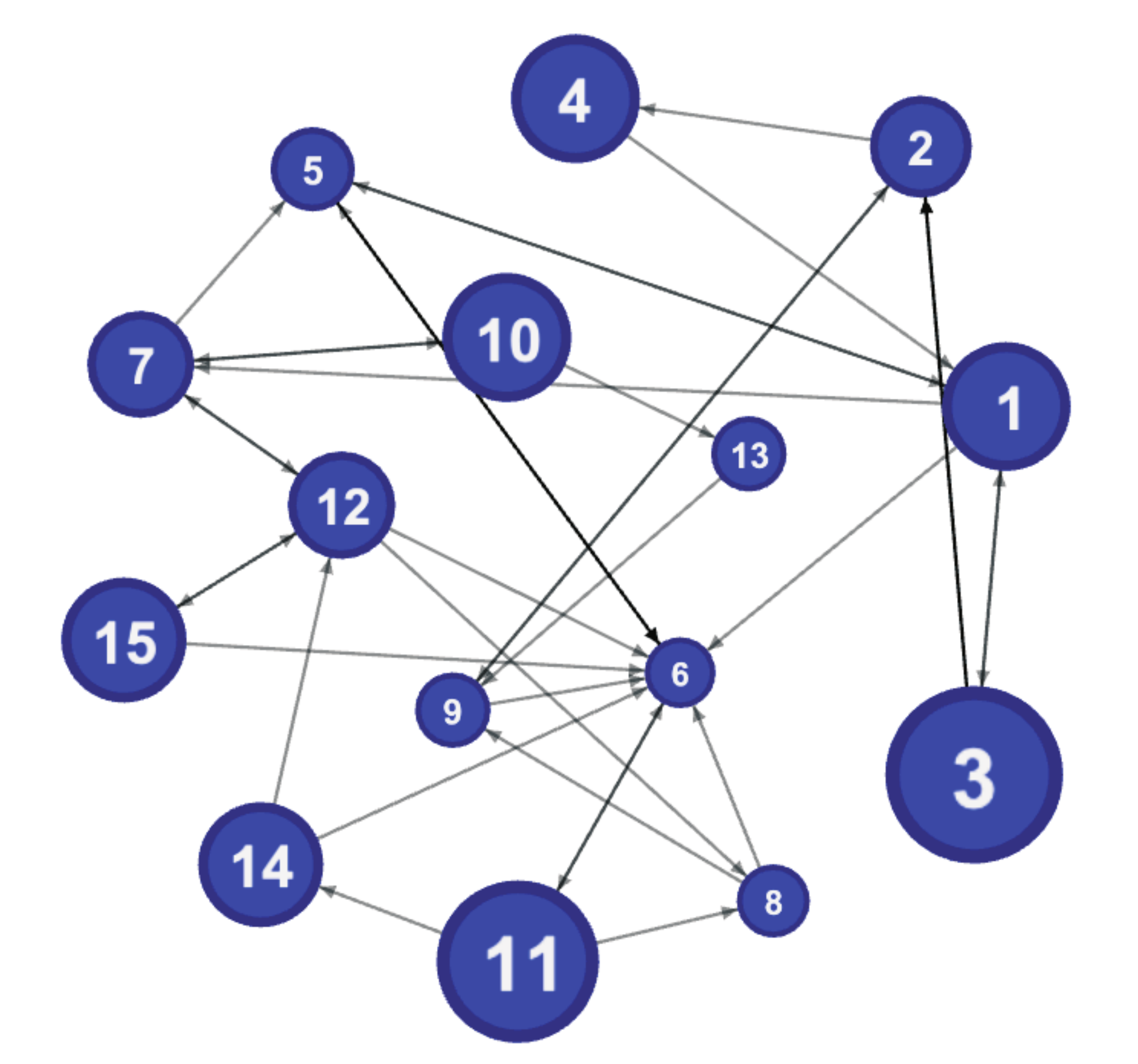}
  \includegraphics[trim=1cm 7cm 1cm 5cm, clip,width=0.55\textwidth]{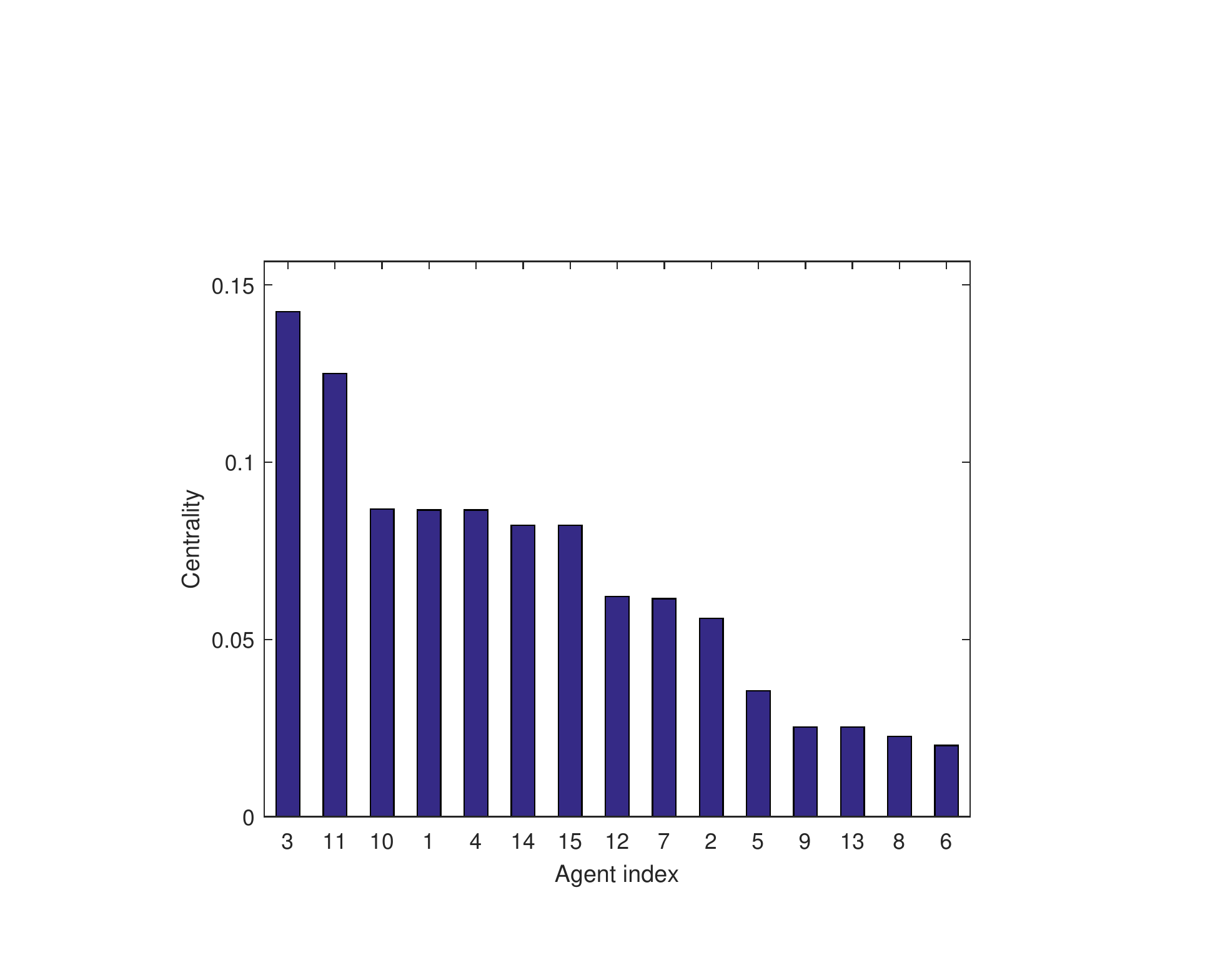}
  \caption{Left: small-scale weakly connected graph. Right: agent centralities, sorted in descending order.}\label{smallscale}
\end{figure}
\begin{figure}[!htb]
  \vspace*{1em}
  \centering
  \includegraphics[trim=1cm 7cm 1cm 5cm, clip,width=0.55\textwidth]{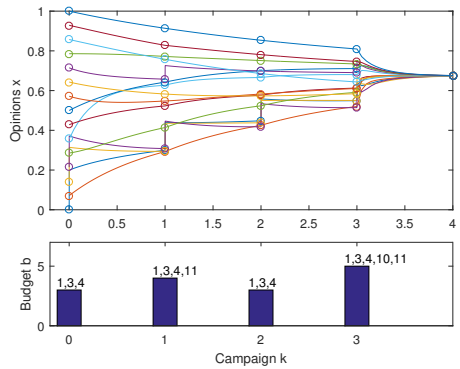}
  \caption{Results for short campaigns. The bottom bar plot shows the budget allocated by the algorithm at each campaign, with the agents influenced in each campaign shown above each bar. The top plot shows the opinions of the agents, with an additional, long campaign converging to the average opinion (so the last campaign duration is not to scale). The circles indicate the opinions \emph{right before} applying the control at each campaign; note the discontinuous transitions of the opinions after control.}\label{shortstage}
 \vspace{-1cm} 
\end{figure}

The small-scale problem has $N=15$ agents, and we start with the weakly connected graph from Figure \ref{smallscale}, left. The target opinion $d$ is $1$. The initial opinions of the agents are distributed on an equidistant grid, starting from $0$ for agent $1$, up to $1$ for agent $15$ (so agents with smaller indices have opinions closer to zero). The centrality of each agent is shown in Figure \ref{smallscale}, right. There are $4$ campaigns, corresponding to $M=3$, and the budget $Q=N=15$ and $\bar{u}=0.2$. For a short campaign length $\delta_k = 0.5\ \forall k$, the brute-force approach gets the results from Figure \ref{shortstage}. The final cost (each individual agent's difference from the desired opinion) is $0.3259$. Examining now the list of agents influenced, we see that these agents are generally among those with large centralities. Nevertheless, relatively lower-centrality agents are preferred when their opinion is far from the target (as is the case for agents $1$ and $4$, whose initial opinion is small). 

To better see the advantages of a well-designed advertising strategy, we compare the results above with the uncontrolled case (no advertising), and to the broadcast strategy (which consists of spending the entire budget at the initial time, with $\bar u$ allocated to each agent). The cost without using any control in this situation is $0.5135$, and the cost with the broadcast strategy is $0.4108$, i.e., we observe a $20$\% gain over the uncontrolled system using the broadcast and around $20$\% gain over the broadcast using the optimal strategy.

\begin{figure}[!htb]
  \vspace*{1em}
  \centering
  \includegraphics[trim=1cm 8cm 1cm 5cm, clip,width=0.55\textwidth]{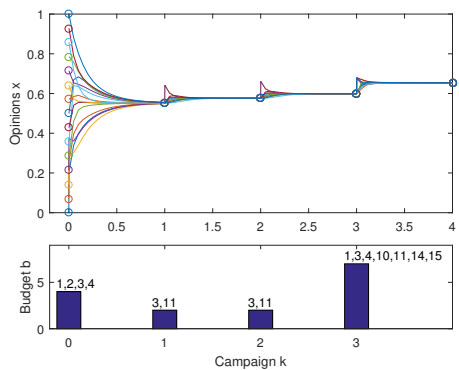}
%   \hfil
%   \raisebox{3cm}{\begin{tabular}{|c|c|}
%     \hline
%     % after \\: \hline or \cline{col1-col2} \cline{col3-col4} ...
%     Campaign & Agents \\ \hline
%     0 & 3,6,7,8 \\\hline 
%     1 & 2,3,9 \\\hline 
%     2 & 2,3,7,9 \\\hline 
%     3 & 2,3,7,9 \\\hline 
%   \end{tabular}
%   }
 \vspace{-0.3cm} 
 \caption{Results for long campaigns. The continuous opinion dynamics is plotted for a sufficiently long time to observe the long campaign behavior, \ie the convergence of opinions of the agents (which means the horizontal axis is not to scale).}\label{longstage_dp}
\end{figure}

\begin{figure}[!htb]
  \vspace*{1em}
  \centering
  \includegraphics[width=0.4\textwidth]{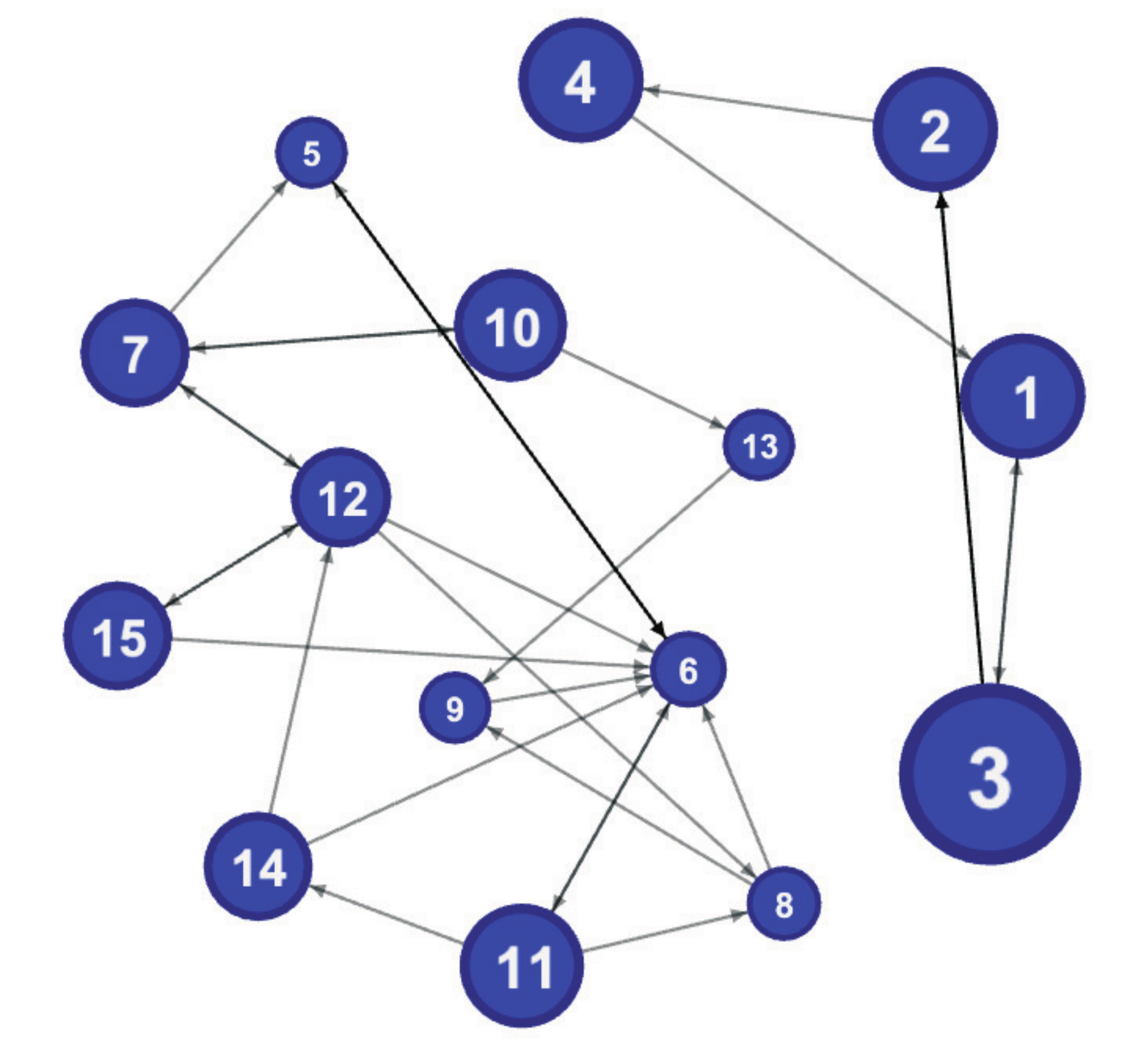}
  \includegraphics[trim=1cm 7cm 1cm 5cm, clip, width=0.55\textwidth]{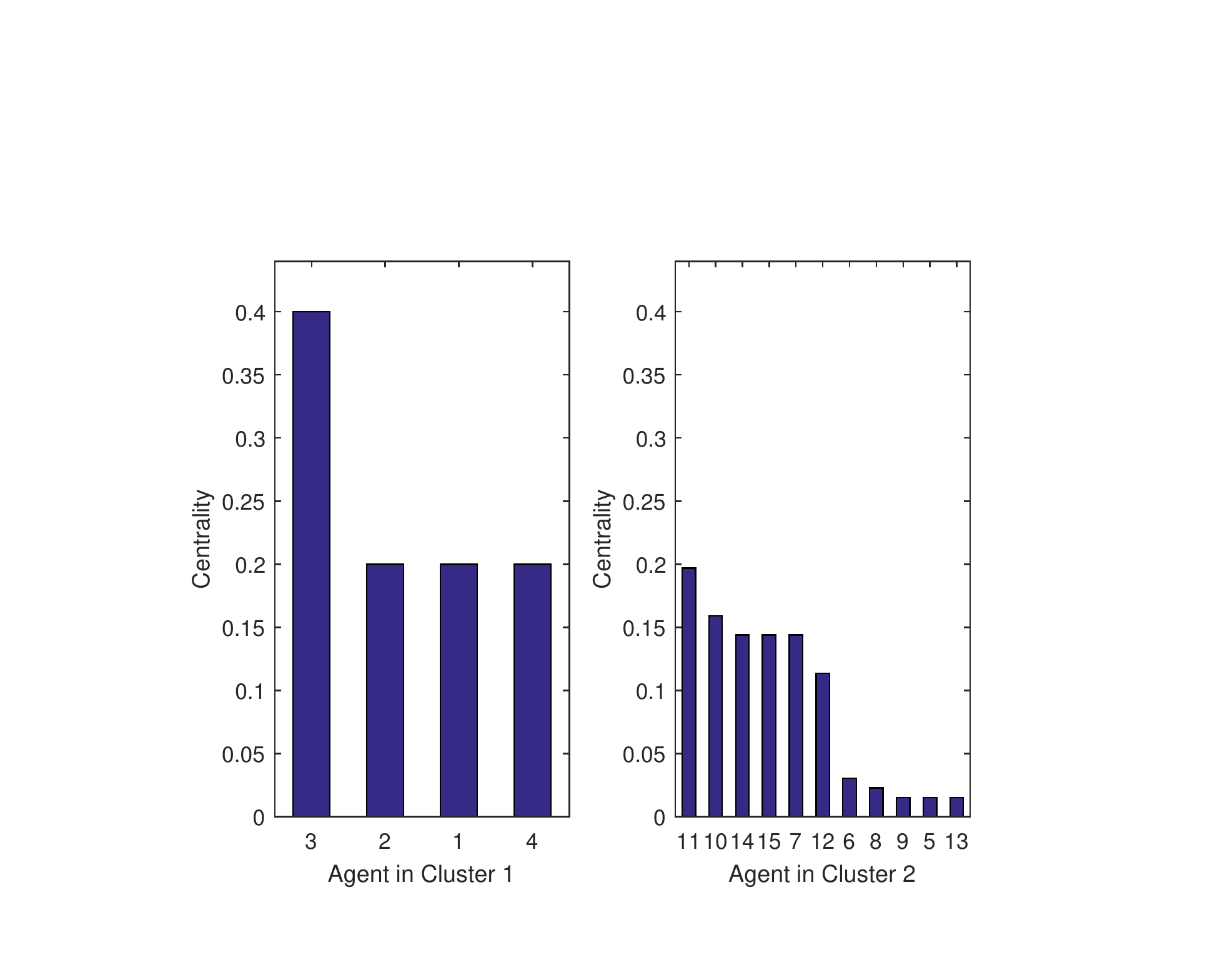}
  \caption{Left: clustered graph. Right: agent centralities in the two clusters.}\label{clustered}
\end{figure}

\begin{figure}[!hbt]
  \vspace*{1em}
  \centering
     \includegraphics[trim=1cm 7cm 1cm 5cm, clip,width=0.55\textwidth]{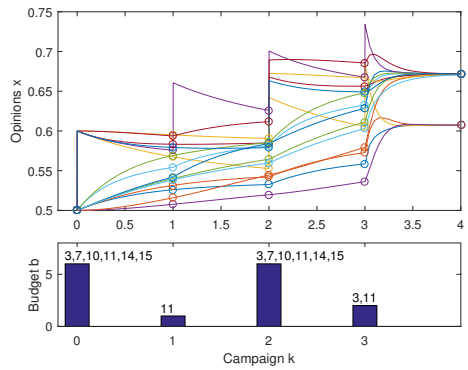}
    % \hfil
    % \raisebox{3cm}{
    %   \begin{tabular}{|c|c|}
    %     \hline
    %     % after \\: \hline or \cline{col1-col2} \cline{col3-col4} ...
    %     Campaign & Agents \\ \hline
    %     0 & 1,3,6,7,8,9,10,12,14 \\\hline
    %     1 & 1 \\\hline
    %     2 &  \\\hline
    %     3 & 1,6,7,10 \\\hline
    %   \end{tabular}
    % }
  \caption{Results for the clustered problem.}\label{fig:clusterresults}
\end{figure}

For the second experiment, in the same network of agents, we consider long campaigns, i.e., $t_{k+1}-t_k \to \infty$. We apply DP, with the results shown in Figure \ref{longstage_dp}. The solution is different from the short-campaign case, which is especially visible at campaigns $k \geq 1$, where only the most central agents are influenced. The final cost is slightly larger, $0.3457$. To better understand the meaning of the long campaigns, note that the network can be (informally speaking) associated with a time constant $T$ equal to the inverse of the smallest real part among all the eigenvalues of the Laplacian $L$ excluding the zero eigenvalue, and as soon as $t_{k+1}-t_k$ is significantly larger than $4T$, the network effectively reaches consensus in-between campaigns so we may consider we are in the long-campaign case. For the particular graph here, $T \approx 3.28$. Note that we can directly compare this long-campaign result with no-control and broadcasting above (since those strategies are independent of campaign length), and we still see significant improvement over both.\\
Next, to illustrate the results of Section \ref{sec:cluster}, we take the graph in Figure \ref{smallscale} and remove all the links between agents $1$ to $4$ and the rest of the agents, obtaining the graph in Figure \ref{clustered}. This new graph has two clusters, the first consisting of agents $1$ to $4$, and the second of the rest of the agents. Four campaigns of length $0.5$ are considered, like before. The brute-force algorithm is applied with $Q=N=15$, starting from an initial state of the network where all agents have opinion $0.5$ (this is done so that they all have the same initial deviation from the desired state, which better exposes the influence of their centrality and group size). The results are shown in Figure \ref{fig:clusterresults}. It is interesting to observe that despite their lower centrality, many agents in cluster $2$ (e.g.\ 7, 10, and 14) are given preference over agents 1, 2, and 4 in cluster $1$. This is because the number of agents is larger in the second cluster, and the selection criterion \eqref{eq:cluster} takes this into account. To compare, we have the final opinion without control to be $0.5$ (since all agents start with the same opinion), with the broadcast strategy to be $0.4$, and with the optimal strategy to be around $0.34$, which corresponds to $85$\% of the cost with the broadcast strategy.

%  0.3839


Finally, we move on to the problem where we test the scalability of DP for large graphs and many campaigns. Specifically, we take $100$ agents and $20$ campaigns. Link weights generated from a uniform distribution over $[0,1]$, after which any link with a weight smaller than $0.3$ is removed. Initial opinions are equidistantly spaced in $[0,1]$ as before, and the total budget is $Q=N=100$. The final cost here is $0.38$ instead of $0.5$ without any control. The obtained cost is close to the one related to the broadcast strategy (which is $0.4$) because all nodes have very similar centrality. Consequently, the DP cost is 76\% of the cost without any control and 95\% of the cost with broadcast strategy. As expected large part of the budget (47\%) is used on the first campaign towards agents having initial condition closer to 0. Note that the brute-force approach would be entirely unfeasible in this problem, while the execution time of DP in Matlab is around 1.7\,s on an Intel i7-3540M CPU.

% \begin{figure}[!htb]
%   \vspace*{1em}
%   \centering
%   \includegraphics[width=0.45\textwidth]{img/vx0_largegraph.eps}
% \caption{The centralities of the most influential few agents in the large graph, together with their initial opinions.}\label{largescale}
% \end{figure}

%\begin{figure}[!htb]
 % \vspace*{1em}
  %\centering
  %\includegraphics[width=0.65\textwidth]{img/longstage_dpmult_large}
%   \hfil
%   \raisebox{3cm}{\begin{tabular}{|c|p{.26\textwidth}|}
%     \hline
%     % after \\: \hline or \cline{col1-col2} \cline{col3-col4} ...
%     Campaign & Agents \\ \hline
%     0 &  
%     2,3,7,13,21,22,24,28,29,35,36, 
%     37,38,39,40,41,42,44,52,53,55, 
%     58,59,60,61,66,68,71,72,73,74, 
%     77,79,80,82,86,87,88,89,90,91, 
%     93,94,97,99\\\hline 
%     1 & 94,97 \\\hline 
%     2 & 94,97 \\\hline 
%     3 to 19 & 94,97,98 \\\hline 
    % 4 & 94,97,98 \\\hline 
    % 5 & 94,97,98 \\\hline 
    % 6 & 94,97,98 \\\hline 
    % 7 & 94,97,98 \\\hline 
    % 8 & 94,97,98 \\\hline 
    % 9 & 94,97,98 \\\hline 
    % 10 & 94,97,98 \\\hline 
    % 11 & 94,97,98 \\\hline 
    % 12 & 94,97,98 \\\hline 
    % 13 & 94,97,98 \\\hline 
    % 14 & 94,97,98 \\\hline 
    % 15 & 94,97,98 \\\hline 
    % 16 & 94,97,98 \\\hline 
    % 17 & 94,97,98 \\\hline 
    % 18 & 94,97,98 \\\hline 
    % 19 & 94,97,98 \\\hline 
%     \end{tabular}
%   }
% \vspace{-0.3cm} 
 % \caption{Results for the large problem. The lists of agents influenced at each campaign are not shown, as they would not be very informative.}\label{longstage_dp_large}
%\vspace{0.3cm} 
%\end{figure}

\section{Conclusions}

In this paper, we have proposed a mathematical formulation of the problem of target marketing over social networks. We show how to exploit some properties of the social network graph in the design of the marketing budget allocation over the agents and over marketing campaigns. A marketer should mainly consider the initial opinion of an agent, its centrality, and (when relevant) the number of agents in the cluster it belongs to. Based on this, we have defined appropriate quantities which measure the influence power of an agent and which allows the marketer to define an order in which it has to allocate its influence budget. The derived budget allocation policies are shown to have a water-filling-type structure. The conducted numerical analysis allows one to extract many precious insights on how to invest a budget over consumers and time. For instance, key consumers to be influenced immediately appear, the number of campaigns to be performed is easily obtained, and the impact of having an advanced marketing campaign (versus allocating the available budget uniformly over consumers and campaigns) can be quantified.

\section*{References}
\bibliographystyle{elsarticle-num}
\bibliography{NAHS}
\end{document}